\documentclass[twocolumn,showpacs,pra]{revtex4}
\usepackage[latin9]{inputenc}
\usepackage{amsmath}
\usepackage{amssymb}
\usepackage{graphicx}
\usepackage{esint}

\usepackage{dcolumn}\usepackage{bm}

\begin{document}

\title{Ionization-induced laser-driven QED cascade in noble gases}

\author{I. I.~Artemenko, I. Yu. Kostyukov}

\email{kost@appl.sci-nnov.ru}

\affiliation{Institute of Applied Physics, Russian Academy of Science, 46 Uljanov
str., 603950 Nizhny Novgorod, Russia}
\begin{abstract}
A formula for the ionization rate in extremely intense electromagnetic
field is proposed and used for numerical study of QED
(quantum-electrodynamical) cascades in noble gases in the
field of two counter-propagating laser pulses. It is shown that the number of the electron-positron pairs produced in the cascade increases with
the atomic number of the gas where the gas density is taken to be reversely proportional to the atomic number. 
While the most electrons produced in the laser pulse front are expelled
by the ponderomotive force from region occupied by the strong laser
field there is a small portion of the electrons staying in the laser
field for a long time until the instance when the laser field is strong
enough for cascading. This mechanism is relevant for all
gases. For high-$Z$  gases there is an additional mechanism associated with
the ionization of inner shells at the the instance
when the laser field is strong enough for cascading. 
The role of both mechanisms for cascade initiation  is revealed.
\end{abstract}

\pacs{12.20.-m,79.70.+q,42.50.Ct,52.27.Ep}

\maketitle

\section{Introduction}

Recently (quantum-electrodynamical) QED cascades in a strong laser
field attracts much attention \cite{Bell2008,Piazza2012,Narozhnyi2015}.
The upcoming laser facilities will be able to generate laser pulses
with the total power up to $10$~PW \cite{eli,apolon}. It is generally
believed that such power can be sufficient to observe QED cascading
in laboratory condition \cite{Gelfer2015,Gonoskov2016,Tamburini2016}.
A cascade develops as a sequence of elementary QED processes: photon
emission by electrons and positrons in the laser field alternates
with pair production as a result of interaction between high energy
photon and laser photons (Breit-Wheeler process \cite{Breit1934}).
Such sequence leads to avalanche-like production of electron-positron
plasma and gammas. The number of the cascade particles can be so great
that they will affect the laser field dynamics. In particularly, the
laser field can be absorbed in self-generated plasma \cite{Fedotov2010,Nerush2011}.

Several configurations of the laser field are proposed to minimize
laser power needed for cascading. One of the most simple configuration
is the superposition of two counter-propagating laser pulses. It is
shown \cite{Jirka2016} that the linear polarization of laser radiation
is more favorable for cascading than circular one in low intensity
limit. The laser-dipole wave can provide development of QED cascade
at the laser power below $10$~PW \cite{Gonoskov2016}. The field
structure which is very similar to the dipole wave can be formed by
$\mbox{12}$ laser pulses \cite{Gonoskov2012}. Another laser configuration
providing QED cascading at power level below $10$~PW can be constructed
by coherent summation of sevaral laser pulses with elliptical polarization
\cite{Gelfer2015}. The focal spot size has a crucial importance for
QED cascading \cite{Tamburini2016}. On the one hand, by reducing
the size of the focal spot at a given power it is possible to increase
the intensity of the laser field, thereby increasing the probability
of QED processes. On the other hand, if the spot size is not large
enough the cascade particles may escape quickly from cascade volume
thereby suppressing cascade development.

In the high intensity limit the cascade can be initiated by the spontaneous
creation of electron-positron pairs out of vacuum (self-seeded QED
cascades) \cite{Fedotov2010}. In the low intensity limit and near
the intensity threshold, the seed particles are needed to trigger
cascading. The seeded particles can be either electrons \cite{Gelfer2015,Nerush2011,Jirka2016}
or high energy photons \cite{nerush-nima}. The electrons as a light
particles can be expelled from the cascade region by the ponderomotive
potential of the laser field before the field strength reaches a maximum
and only a small portion of the seed electrons may survive to trigger cascade \cite{Jirka2016}.
Expulsion of the highly relativistic electrons by the ponderomotive
force is suppressed due to the relativistic gain in the electron mass
\cite{Mironov2016}. Yet the use of relativistic electrons as seed
particles is hindered by high cost of high energy electron accelerators.
In addition this also requires focusing of the electrons to the interaction
region and synchronization between the electron beam and laser pulses.
The same reasons (high cost of bright gamma ray sources, focusing
and synchronization of the gamma beam) may prevent the use of high
energy photons as seed particles. 

Gases with high-Z atoms can be a source of the seed electrons. The
ionization potential of the inner electrons of high-Z atoms can be
so large that such electrons can leave the atoms at very high laser
field strength. Therefore the seed electrons can be produced by field
ionization when the laser field strength peaks and is strong enough
for cascading. It is demonstrated recently \cite{Tamburini2016}
that cascade triggering in the field of two counter-propagating may be
facilitated by employing suitable high-Z gases. However the simplified
model for atom ionization was used and only hydrogen and oxygen are explored for gas target. The model does not take into account
the probabilistic nature of ionization, dependence of the ionization
probability on the shell electron parameters, sequential and multiple
ionization of high-Z atoms. As a result this model cannot provide
accurate description of the ionization and the dynamics of the seed
electrons. In our work QED cascading in all noble gases irradiated
by counter-propagating laser pulses is studied by three dimensional particle-in-cell
Monte Carlo (3D PIC-MC) simulations with more realistic approach to
laser ionization. We propose a new ionization rate formula that extended
the known formula for tunnel ionization \cite{Perelomov1966-1,Popov2004}
to extremely intense field when the potential barrier are strongly
suppressed.

It should be noted that the foils made from high-Z material can be
also used as a laser target and the source of the seeded electrons
\cite{Artemenko2016}. However because of large target density QED
cascade development can be affected by collisional processes like
bremsstrahlung and electron-positron pair production as a result of
photon scattering by nuclei. Here we discusses the use of rarified
gases in order to neglect collisional processes. 

The paper is organized as follows. The field ionization model is described
in Sec. II. In Section III, the results of 3D PIC simulations of QED
cascades are presented. The distribution
and the spectrum of the cascade particles are calculated. Section~VI
contains discussion and conclusions. The contribution of collisional effects is estimated and discussed.

\section{Ionization model }

Effect of a strong electromagnetic field on atom may lead to ionization.
In the tunnel regime of ionization the atom electrons penetrate through
the potential barrier formed by the atomic field and the external electric
field. At low intensities the field ionization occurs in multiphoton
regime. The regime of the field ionization depends on the Keldysh
parameter $\gamma_{K}= a^{-1} \left( 2 I_i/ m_e c^2 \right)^{1/2} $, where
$I_{i}$ is the ionization potential of the ion, $a=eE_{L}/\left(m_{e}c\omega_{L}\right)$
is the dimensionless laser field, $E_{L}$ and $\omega_{L}$ are the
laser field strength and the laser frequency, respectively, $e$ and
$m_{e}$ are the charge and mass of the electron, respectively, $c$ is the speed of light \cite{Keldysh}. It is generally believed
 that the field ionization occurs in the tunnel regime
if $\gamma\lesssim0.5$ \cite{Ilkov}. In our simulations the electromagnetic field
can be treated as a static within the code time step. The rate of ionization
 in the static electric field is in the tunnel regime \cite{Perelomov1966-1,Popov2004,Karnakov2015}:

\begin{eqnarray}
\nonumber
W_{TI} & = & \omega_a  \kappa^2 C_{kl}^2 \left(\frac{2}{F}\right)^{2n^{*}-m-1}
\\ 
&\times& \frac{(l+m)!(2l+1)}{2^{m}m!(l-m)!}\cdot\exp\left(-\frac{2}{3F}\right),
\label{Wppt} \\ \nonumber
C_{kl}^{2} & = & \frac{2^{2n^{*}}}{n^{*}\Gamma(n^{*}+l^{*}+1)\Gamma(n^{*}-l^{*})},
\end{eqnarray}
where $F=E/\left(\kappa^{3}E_{a}\right)$ is the normalized electric
field,  $n^{*}=Z/\kappa$ is the effective
principal quantum number of the ion, $Z$ is the ion charge number, $\kappa^{2}=I_{i}/I_{H}$, $I_{H}=m_{e}e^{4}/\left(2\hbar^{2}\right)\simeq13.59843$eV is the ionization potential of hydrogen,
$l^{*}=n^{*}-1$  is the effective angular momentum, $l$ and
$m$ are the orbital and magnetic quantum numbers, respectively,  $E_{a}=m_{e}^{2}e^{5}\hbar^{-4}\approx5.14224\cdot10^{9}$V/cm
is the atomic electric field $\omega_{a}=m_{e}e^{4}\hbar^{-3}\simeq4.13\cdot10^{16}\mbox{c}^{-1}$
is the atomic frequency, $\hbar$ is the
Planck constant, $\Gamma(x)$ is the Gamma function \cite{Abramowitz}.
In the limit $n^{*}\gg1$ formula~(\ref{Wppt}) reduces to the ionization
rate given in Ref.~\cite{Ammosov1986}. 

The formula~(\ref{Wppt}) is valid when the unperturbed atomic energy
level is much lower then the potential barrier maximum. This condition is fulfilled when the
external field strength is the much less than the critical field $E\ll E_{cr}=E_{a}\kappa^{4}/\left(16Z\right)$
\cite{krainov1998}. For hydrogen-like atoms and ions with regard
to the Stark effect $E_{cr,H}=\left(2^{1/2}-1\right)E_{a}$ \cite{Bauer1997}.
We will use expression $E_{cr}=E_{a}\kappa^{4}/\left(16Z\right)$
because $E_{cr}<E_{cr,H}$ and the ionization rate given by $W_{TI}$
strongly deviates from results of numerical simulations for $E>E_{cr}$
\cite{Tong2005}. 

The barrier suppression regime of the field ionization is relevant
if $E>E_{cr}$. The analytical description of this regime is difficult
since the perturbation methods are no longer valid for $E\sim E_{cr}$.
For example, the analytical formula derived in Ref.~\cite{Krainov1997}
for the field ionization rates in the barrier suppression regime does
not agree with numerical time-dependent Schredinger equation (TDSE)
calculations for $E>E_{cr} $\cite{Bauer1999}. 

Several empirical formulas based on
numerical simulations have been proposed for the ionization rate in
the tunnel and barrier suppression regimes \cite{Tong2005,Bauer1999,Zhang2014}.
In Ref.~\cite{Bauer1999} the piecewise formula for ionization rate
is proposed so that $W_{TI}$ is used for $E \leq E_{TIQ}\sim E_{cr}$ while
the quadratic dependence of the rate on the field strength is assumed
for $E>E_{TIQ}$ 

\begin{eqnarray}
W_{Q}(E) & = & \omega_{a}2.4(E/E_{a})^{2},\label{Wq}
\end{eqnarray}
where $E_{TIQ}$ is a threshold electric field determined by imposing
$W(E)$ to be continuous $W_{TI}(E_{TIQ})=W_{Q}(E_{TIQ})$. Other
empirical formulas providing continuous transition between tunnel and
barrier suppression regimes are presented in Refs.~\cite{Tong2005,Zhang2014}. 

\begin{figure}[h]
\centering \includegraphics[width=8cm]{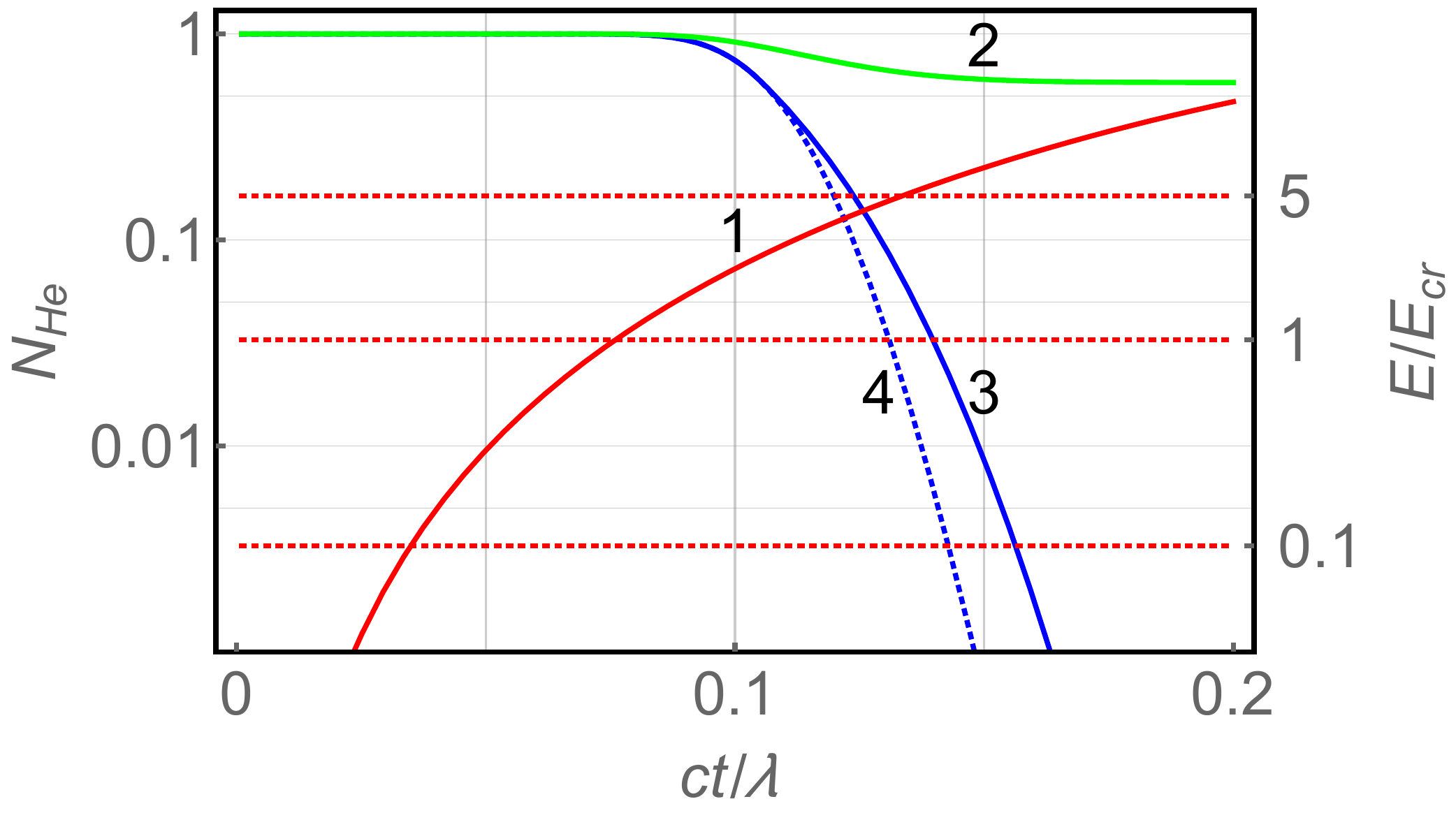}

\caption{The dependence of the laser field strength on time (red line
1) and the probabilities of He to be not ionized as a function of
time (lines 2-4). The probabilities are calculated by integrating
$W_{TI}$ (dotted blue line 4),  $W_{PW}$ (solid blue line 3) and
 the ionization rate proposed in Ref.~\cite{Tong2005} (green line 2) over time. }
\label{heionization} 
\end{figure}

The proposed formulas for the ionization rate become not accurate in the limit of
the extremely high field. For example according to numerical TDSE
simulations the dependence of the ionization rate on the field strength
is close to linear rather than to quadratic for $E>0.4E_{a}\gg E_{cr}$
\cite{Bauer1999}. The formula proposed in Ref.~\cite{Tong2005}
predicts reduction of the ionization rate in the limit $E\gg E_{cr}$
that does not agree with numerical simulations \cite{Zhang2014}. However
the ionization rate formula, which is valid for $E\gg E_{cr}$, is
needed to analyze field ionization for laser intensity above $10^{23}$~W/cm$^{2}$
when QED cascading is possible. For example, Eq.~(\ref{Wppt}) for
tunnel ionization predicts that $90\%$ ionization of He in the electric field

\begin{eqnarray}
E(t) & = & a\frac{mc\omega_{L}}{e}\sin\left(\omega_{L}t\right)\sin^{2}\left(\frac{t}{T}\right)\label{Et1}
\end{eqnarray}
occurs when the laser field strength achieves value $E\approx4E_{cr}$
(see Fig.~\ref{heionization}), where $0 \leq t \leq  20 \lambda /c $, $a=500$, $\omega_{L}=2\pi c/\lambda$,
$T=40\omega_L c$, $\lambda=1$~$\mu$m is the laser wavelength. The
probability for He to be not ionized can be calculated as follows
$P(t)=1-\exp\left\{ -\int_{-\infty}^{t}W_{TI}[\left|E\left(\tau\right)\right|]d\tau\right\} $.
The ionization rate given by Eq.~(\ref{Wppt}) is in orders of magnitudes
greater than that numerically calculated for $E\gg E_{cr}$  \cite{Tong2005,Bauer1999}. 
Therefore the $90\%$ ionization of He will be
achieved even at higher fields than $4E_{cr}$.

In the limit of extremely strong laser field the ion field can be
neglected and the electrons inside the ion can be considered as unbound
just after the laser field is turned on quickly. The field of the
ion with charge $Z$ at the position of the outer electron can be
estimated as $E_{i}(r_{0})\simeq eZ/r_{0}^{2}=16E_{cr}$ at the beginning
of ionization, where $r_{0}\simeq a_{B}Z/\kappa^{2}=4r_{m}(E_{cr})$
is the orbit radius of the electron with ionization potential $I_{i}$,
$r_{m}(E)=\left(eZ/E\right)^{1/2}$ is the position of the potential
barrier maximum for the electron in ion field and the external electric
field $E$ and $a_{B}=\hbar^{2}/(m_{e}c^{2})$ is the Bohr radius.
Therefore the ion field can be neglected if $E\gg16E_{cr}$. However
$16E_{cr}$ is the maximal value of the ion field that electron feels
during ionization. The condition for neglecting of the ion
field can be taken as $E\gg E_{cr}$ because in this limit the ionization
energy is much higher than the potential barrier maximum. 

The ionization time can be estimated from model of a free electron
as the time needed to accelerate the electron from the energy of the
atomic level $\varepsilon_{e}=-I_{i}$ to the continuum $\varepsilon_{e}=0$
so that $I_{i}=m_{e}c^{2}\left[\left(1+a^{2}\omega_{L}^{2}\tau_{i}^{2}\right)^{1/2}-1\right]$,
where $\varepsilon_{e}$ is the electron energy, $\tau_{i}$ is the
ionization time and $amc\omega_{L}/e = E$ is the external electric field accelerating
electron. Therefore the ionization rate in the limit of extremely
strong field can be estimated as follows $W_{L}\approx\tau_{i}^{-1}=\omega_{L}a\left[\left(1+I_{i}/m_{e}c^{2}\right)^{2}-1\right]^{-1/2}$.
Neglecting the relativistic corrections ($I_{i}\ll m_{e}c^{2}$) we
get linear dependence of the rate on the electric field

\begin{eqnarray}
W_{L}(E) & \approx & \omega_{L}a\sqrt{\frac{m_{e}c^{2}}{2I_{i}}}=\omega_{a}\frac{E}{E_{a}}\sqrt{\frac{I_{H}}{I_{i}}}\label{fe}
\end{eqnarray}
Finally, making of use piecewise approach, the formula for the field
ionization rate can be extended to the limit of extremely strong field
when the potential barrier is strongly suppressed

\begin{equation}
W_{PW}\left(E\right)=\begin{cases}
W_{TI}(E), & \mbox{if }E\leq E_{TIL},\\
W_{L}(E), & \mbox{if }E>E_{TIL},
\end{cases}\label{general}
\end{equation}
where value $E_{TIL}$ is a threshold electric field and a solution
of the transcendent equation $W_{TI}(E)=W_{L}(E)$. 

\begin{figure}[h]
\includegraphics[width=8cm]{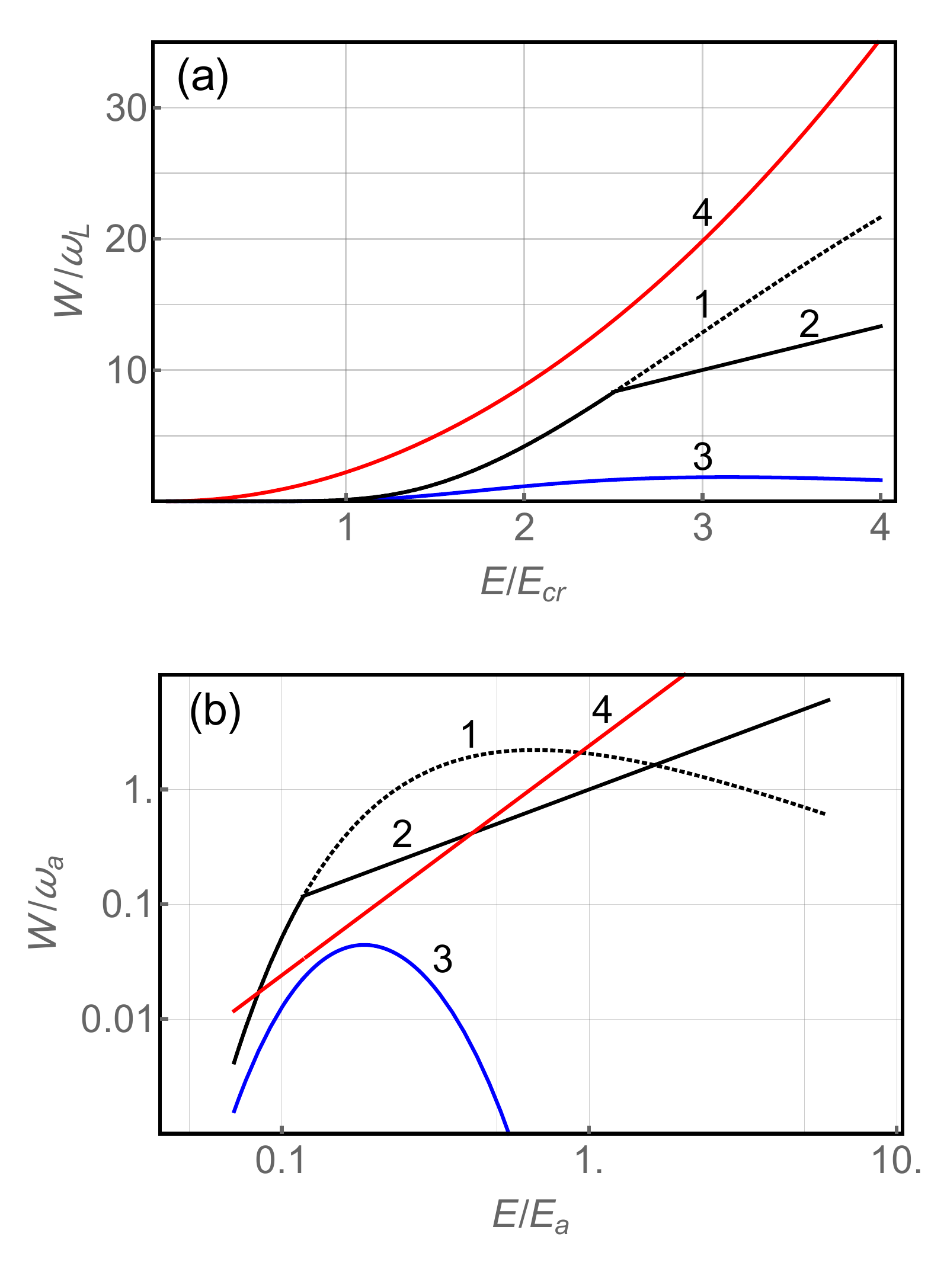}
\caption{(a) The ionization rates $W_{TI}(E) $ (dotted black line 1), $W_{PW}(E) $ (solid black line 2), $W_Q$ (red line 4) and the rate proposed in Ref.~\cite{Tong2005} (blue line 3) as functions of the field strength for single-electron
ionization of He. (b) The ionization rates $W_{TI}(E) $ (dotted black line 1), $W_{PW}(E) $ (solid black line 2), $W_Q$ (red line 4) and the rate proposed in Ref.~\cite{Tong2005} (blue line 3) as functions of the field strength for ionization of hydrogen.}
\label{rate} 
\end{figure}

It is interesting to note that according to our calculation the intersection
between $W_{TI}(E)$ and $W_{L}(E)$ occurs at reasonable field strength
$E_{TIL}\sim E_{cr}$ for all noble gases. For example for all $54$
electrons of Xe $1.15E_{cr}<E_{TIL}<1.91E_{cr}$. If the last two
electrons ($1s^{1}\text{and }1s^{2}$ electrons) of Xe are excluded
then $1.15E_{cr}<E{}_{TIL}<1.45E_{cr}$. The formula proposed in Ref.~\cite{Bauer1999}
as a combination of $W_{TI}(E)$ and $W_{Q}(E)$ predicts unphysical
value of the threshold electric field $E_{TIQ}=0$ for ionization of
$1s^{2}$ electron of He (see Fig.~\ref{rate} (a)). It is worth
to note that $W_{Q}(E)$ for hydrogen starts to significantly deviate
from numerical results at $E=E_{QL}\simeq0.4E_{a}$ where $W_{Q}(E)$
crosses with $W_{L}(E)$ (see Fig.~6 in Ref.~\cite{Bauer1999} and
Fig.~\ref{rate} (b)). 

In order to take into account the multiple ionization within one time
step of PIC code the MC kinetic numerical model is used \cite{Nuter2011}.
The method is based on solution of set coupled first-order differential
equations describing evolution of ion charge state \cite{Rae1992}.
The equations can be solved numerically \cite{Chen2013,Korzhimanov2013}
or analytically \cite{Nuter2011} assuming that the field distribution
and the ionization probabilities do not change within the time step.
Ionization events are modeled by MC numerical scheme as a random process
where the ionization rate is determined by Eq.~(\ref{general}).
The energy losses because of ionization are neglected as they are
much less than the losses associated with QED cascading (see Sec.~IV).

\section{Numerical simulations}

Development of laser-assisted QED cascade in noble gases is studied
by 3D PIC-MC simulations with code QUILL \cite{quill,Nerush2011}. The code part based on PIC method models
dynamics of a plasma and laser field while the part based on MC method
models emission of high energy photons, electron-positron pair creation
and field ionization of atoms and ions. 

\begin{figure}[h]
\centering \includegraphics[width=8cm]{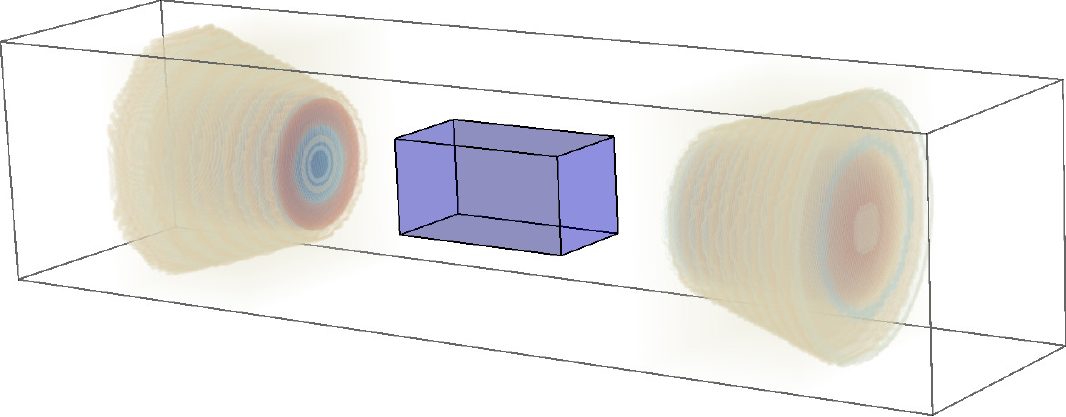}
\caption{The scheme of the laser pulse interaction with gas volume. Two counter-propagating
laser pulses are focused to the gas volume center.}
\label{scheme} 
\end{figure}

In our simulations two laser pulses propagate towards each other along
$x$-axis (see Fig.~\ref{scheme}). The laser pulse centers are located
in the points $x_{0}=16$$\lambda$, $y_{0}=z_{0}=16.5$$\lambda$ and $x_{0}=40$$\lambda$, $y_{0}=z_{0}=16.5$$\lambda$ at $t=0$, respectively, 
where $\lambda=1\mu m$. The pulses are focused on the point $x_{c}=28$$\lambda$,
which is the center of the gas volume. The laser pulses have linear
polarization ($E_z=B_y=0$) and the $y$-component of the electric field at $t=0$
is 
\begin{eqnarray}
E_{y}(\mathbf{r}) & = &A(\mathbf{r})\cos^{2}\left[\frac{\pi\sqrt{y^{2}+z^{2}}}{2\sigma(\mathbf{r})}\right]\varUpsilon\left(\mathbf{r}\right),
\\ 
\nonumber
\varUpsilon\left(\mathbf{r}\right) & = & \cos^{2}\frac{\pi x}{2\sigma_{x}}\cos\psi(\mathbf{r})-\frac{\lambda}{4\sigma_{x}}\sin\psi(\mathbf{r})\sin\frac{\pi x_{s}}{\sigma_{x}},
\\
\nonumber
\psi(\mathbf{r}) & = & R(\mathbf{r})-\arctan\frac{d}{x_{R}}-\arctan\frac{x_{s}-d}{x_{R}},
\\
\nonumber
R(\mathbf{r}) & = & k_{L}\left[x_{s}+\frac{\left(y^{2}+z^{2}\right)\left(x_{s}-d\right)}{2(x_{s}-d)^{2}+2x_{R}^{2}}\right],
\\
\nonumber
x_{s} & = & x-x_{0},
\\
\nonumber
\sigma(\mathbf{r}) & = & \frac{\pi \sigma_{0}}{\sqrt{2^{-2}3\pi^{2}-4}}
\\
\nonumber
&\times& \left[1+\left(\frac{\psi(\mathbf{r})-k_{L}d+\arctan x_{R}^{-1}d}{k_{L}x_{R}}\right)^{2}\right]^{1/2},
\\
\nonumber
A(\mathbf{r}) & = & \frac{amc\omega_{L}}{e}\frac{\sqrt{x_{R}^{2}+d^{2}}}{\sqrt{x_{R}^{2}+x_{0}^{2}}}\frac{\sigma_{0}}{\sigma(\mathbf{r})}\frac{\pi}{\sqrt{2^{-2}3 \pi^{2} -4}},
\end{eqnarray}
where $a=500$ is the laser pulse amplitude, $\sigma_{x}=8\sqrt{2\pi}\lambda$
is the pulse length, $d=12\lambda$ is the distance from the center
of the last pulse to gas volume center, $x_{R}=\pi\sigma_{0}^{2}/\lambda$,
$\sigma_{0}=3\lambda$, $k_{L}=2\pi/\lambda$. The other components
of the electric and magnetic fields at $t=0$ can be calculated from the
Maxwell's equations $\nabla\cdot\mathbf{E}=\nabla\cdot\mathbf{B}=0$.

The gas density is chosen less than $10^{16}$cm $^{-3}$ so that the
collisional effects (collisional ionization, bremsstrahlung, pair
photoproduction by nuclei etc.) can be neglected (see discussion in
Sec.~IV). The seed electrons for cascade triggering
are produced by the field ionization of the gas atoms. Noble
gases He, Ne, Ar, Kr and Xe are explored. In order to study the contribution of the electrons
bound in the different atom shells  the densities
of the gases is chosen to be reversely  proportional to the atomic numbers
so that the number of electrons produced after full ionization  are
the same for all gases. For example, the density of He is $9.03\times10^{15}\mbox{c}\mbox{m}^{-3}$
in our simulations that is in $27$ times higher than the density
of Xe, $3.35\times10^{14}\mbox{c}\mbox{m}^{-3}$. Therefore, in
the case of full atom ionization the densities of the ionization-produced
electrons for both gases are the same.

\begin{figure}[h]
\includegraphics[width=8cm]{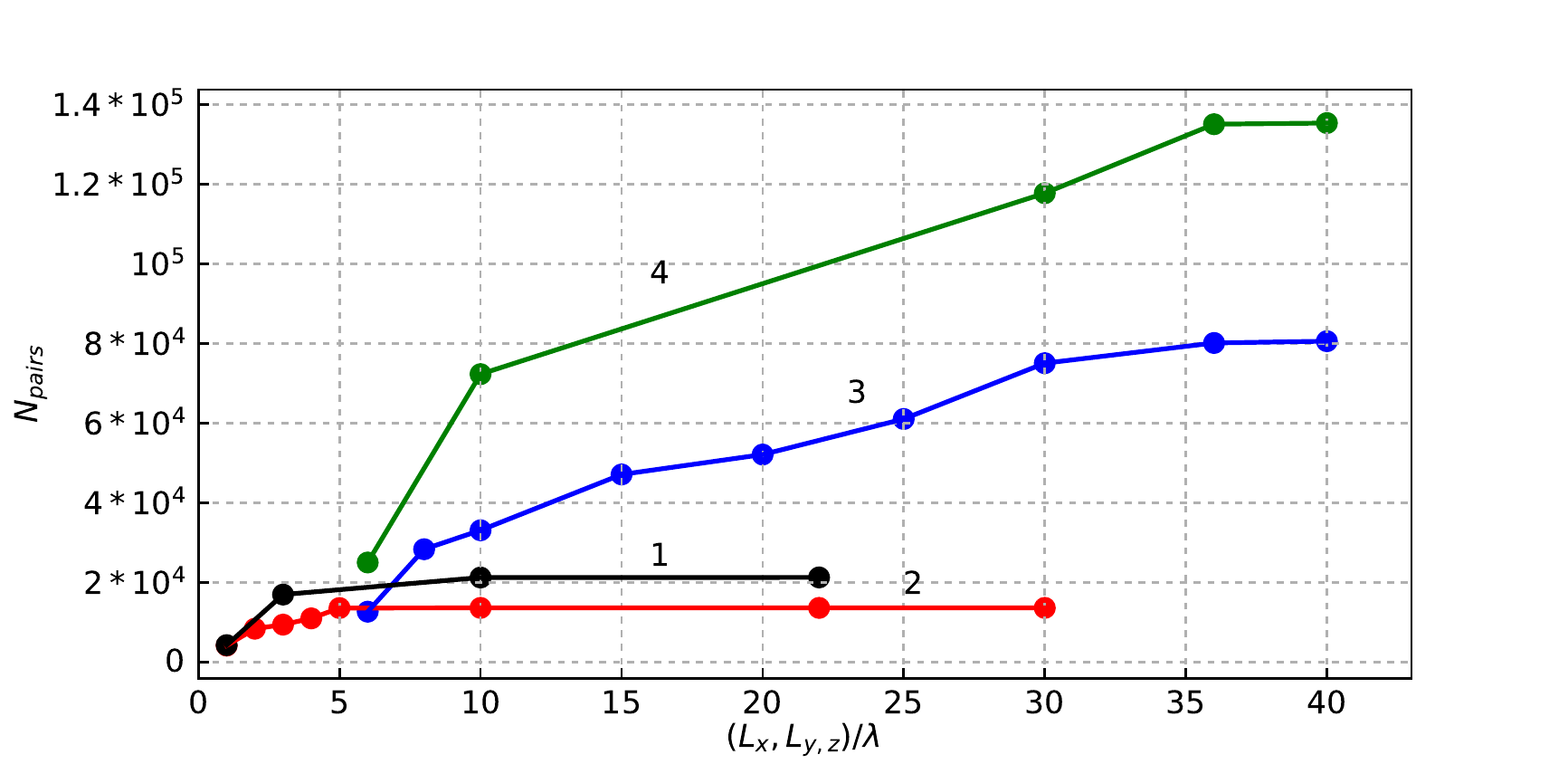}
\caption{The number of the pairs produced in the cascade as a function of the gas volume
length (along $x$-axis) for He (blue line 1) and for Xe (green line
2) as well as a function of the gas volume width (along $y$ and $z$
axises) for He (red line 3) and for Xe (black line 4). }
\label{len} 
\end{figure}

First we study QED cascade development in He. The gas volume
in simulation has a length $40\lambda$ along $x$-axis ($8\lambda\leq x\leq48\lambda$)
and $5\lambda$ along $y$-axis ($14\lambda\leq y\leq19\lambda$)
and $z$-axis ($14\lambda\leq z\leq19\lambda$). Further enlargement
of the gas volume in all directions does not increase the number of
pairs produced in the cascades (see Fig.~\ref{len}). The gas density
is $9.03\times10^{15}\mbox{c}\mbox{m}^{-3}$.

\begin{figure}[h]
\includegraphics[width=8cm]{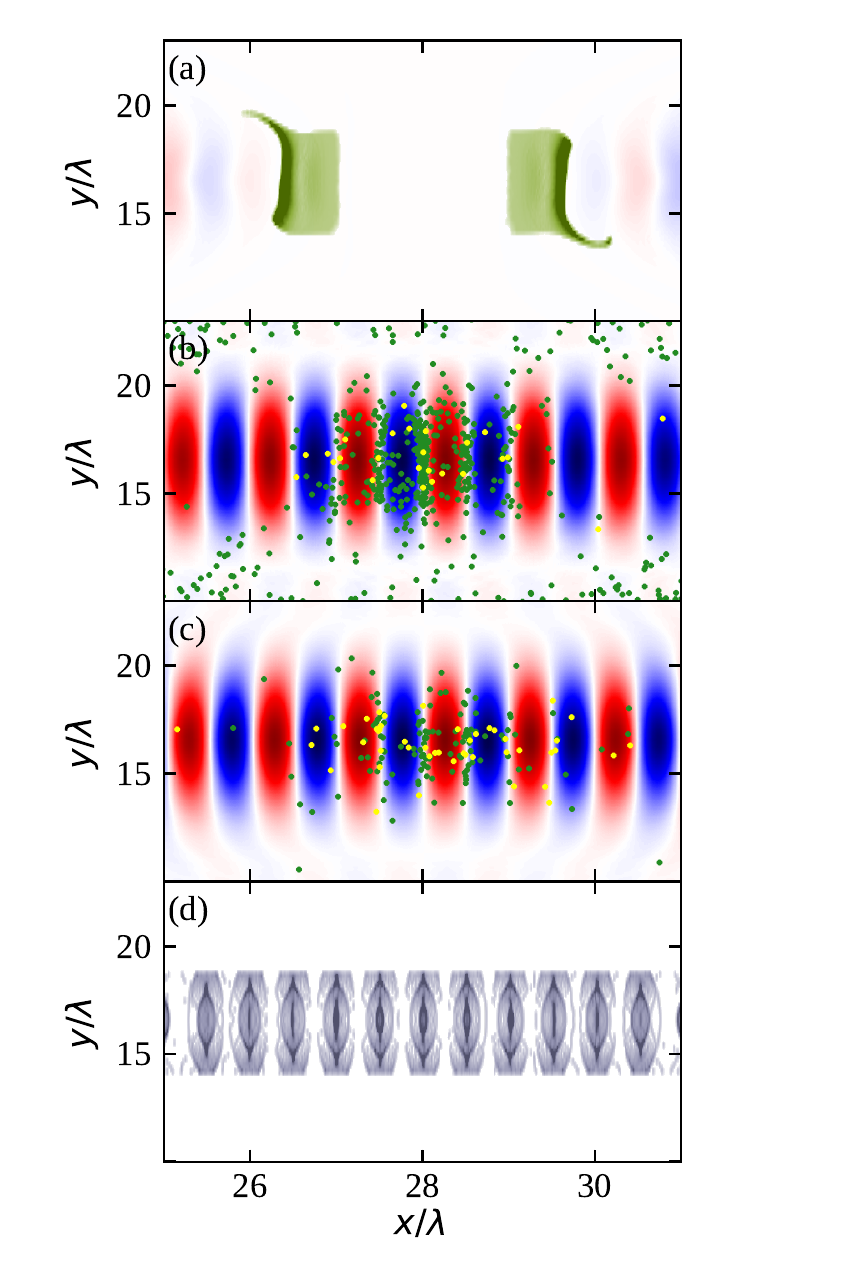}
\caption{The distribution of the electrons (green dots), the positrons (yellow
dots) and the laser field component $E_{y}$ (blue and red) in $x-y$
plane at  $t=8\lambda/c$ (a), $t=15\lambda/c$ (b) and $t=18\lambda/c$
(c) for He. The distribution of the He ions in $x-y$ plane
at $t=18\lambda/c$ (d).}
\label{He-distr} 
\end{figure}

The distributions of the electrons, positrons, ions and $E_{y}$ are
shown in Fig.~\ref{He-distr} in the different moments of time. The
pulse centers cross each other in $x=28\lambda$ at $t=18\lambda/c$.
The counter-propagating laser pulses generate field structure which
is close to the linearly polarized standing wave near $x=28\lambda$.
In the case of He the full ionization of atoms occurs already
at the laser pulse front. For $t\geq10\lambda/c$ the gas is fully
ionized and new electrons are not produced due to field ionization.
Most of the produced electrons are pushed out by ponderomotive force
of the laser pulse from high intensity region in transverse direction
and cannot initiate cascade. The small part of the electrons moves
along with the laser pulses thereby forming two counter-propagating
relativistic bunches (see Fig.~\ref{He-distr}(a)). The motion of
each bunch are stopped by the counter-propagating laser pulse. Moreover
the bunch electrons are trapped in the standing wave nodes corresponding
to the minimum of the ponderomotive potential (see Fig.~\ref{He-distr}(b))
\cite{Jirka2016,Lehmann2012,Gonoskov2014}. 

\begin{figure}[h]
\centering \includegraphics[width=8cm]{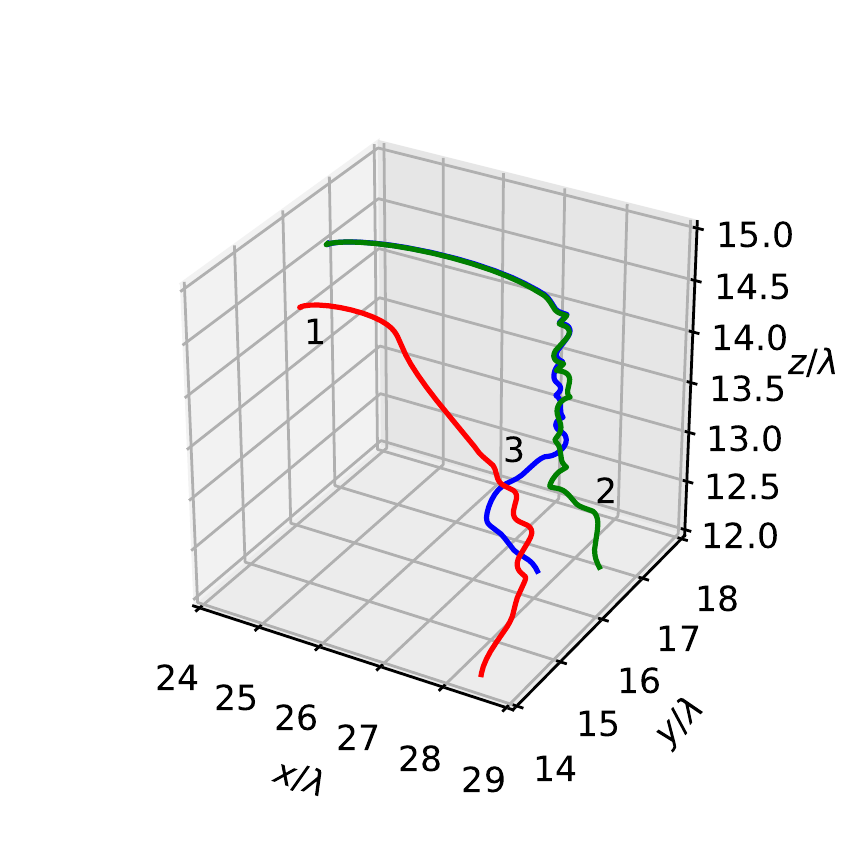} 
\caption{The trajectory of the escaping electrons pushed out by the ponderomotive
force from high intensity region (red line 1) and the trajectory of
the trapped electrons staying for a long time in region, where laser
field peaks, (blue line 2). The electrons are created by ionization
of He atoms.}
\label{He-track} 
\end{figure}

The typical trajectories of the trapped electrons staying for a long
time in region, where laser field peaks ($x=28\lambda$, $y=16.5\lambda$,
$z=16.5\lambda$), and the escaping electrons pushed out by the ponderomotive
force from high intensity region are shown in Fig.~\ref{He-track}.
When the wave strength becomes strong enough the bunch electrons start
to initiate cascade with prolific pair
production. It follows from simulations that the pairs are efficiently
produced within time interval $12\lambda/c\leq t\leq23\lambda/c$.
The ion density also peaks in the wave nodes because of large
uncompensated electron charge accumulated there (see Fig.~\ref{He-distr}(d)).

\begin{figure}[h]
\centering \includegraphics[width=8cm]{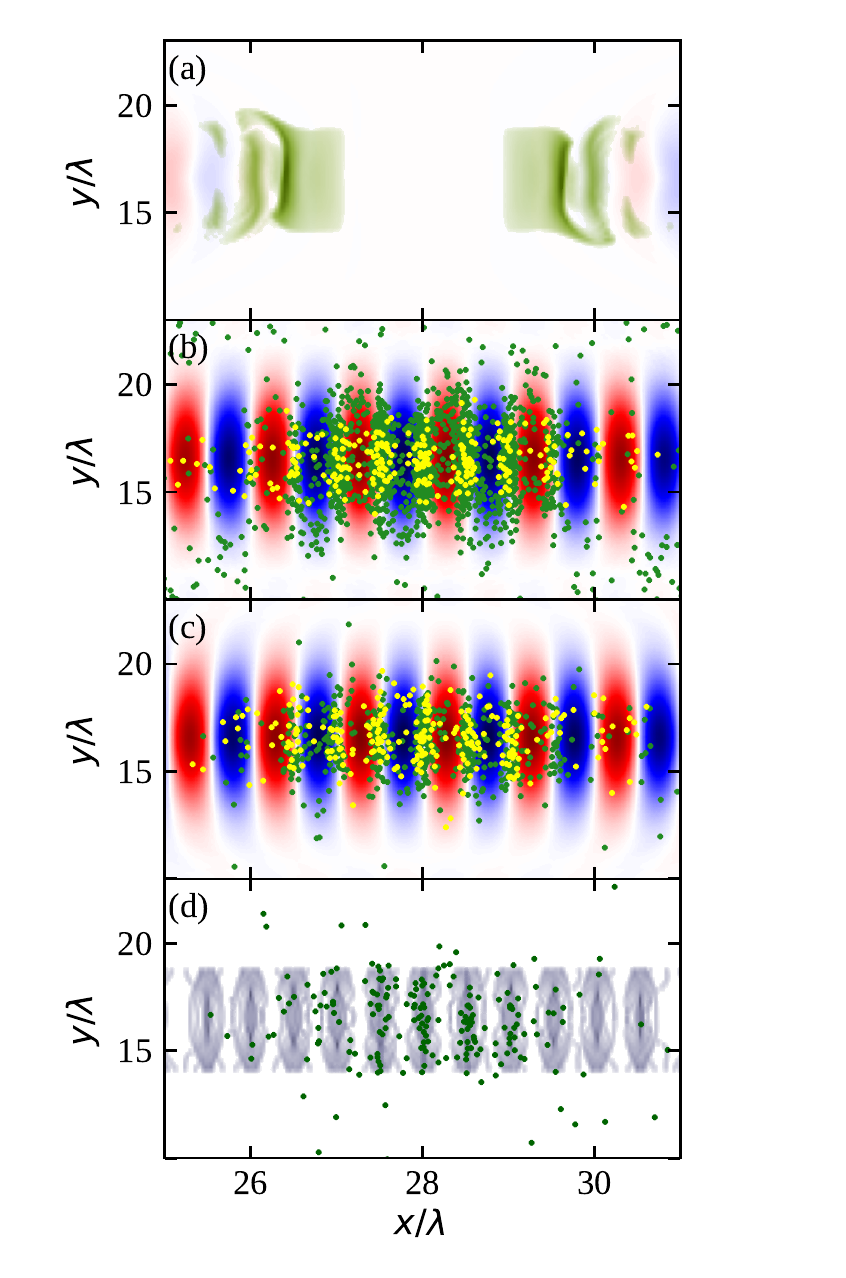} \caption{The distribution of the electrons (green dots), the positrons (yellow
dots) and the laser field component $E_{y}$ (blue and red) in $x-y$
plane at $t=8\lambda/c$ (a), $t=15\lambda/c$ (b) and $t=18\lambda/c$
(c) for Xe. The distribution of the inner-shell ($2s^{1}$) electrons
(green dots) and Xe ions (grey color) in $x-y$ plane at $t=18\lambda/c$ (d).}
\label{Xe-distr} 
\end{figure}

The density of Xe in the simulations is $3.35\times10^{14}\mbox{c}\mbox{m}^{-3}$
that is in $27$ times less than the density of He so that the number of the atomic electrons in the gas volume are the same for both gases.  The cascade development in
Xe is shown in Fig.~\ref{He-track}. Like for He, the small
portion of electrons produced from outer shell of Xe atoms in the
laser pulse fronts forms counter-propagating bunches (see Fig.~\ref{Xe-distr}(a)).
The bunch electrons are trapped in the nodes of the standing wave
generated near $x=28\lambda$, where the laser pulse centers cross
each other (see Fig.~\ref{Xe-distr}(b)).

\begin{figure}[h]
\includegraphics[width=8cm]{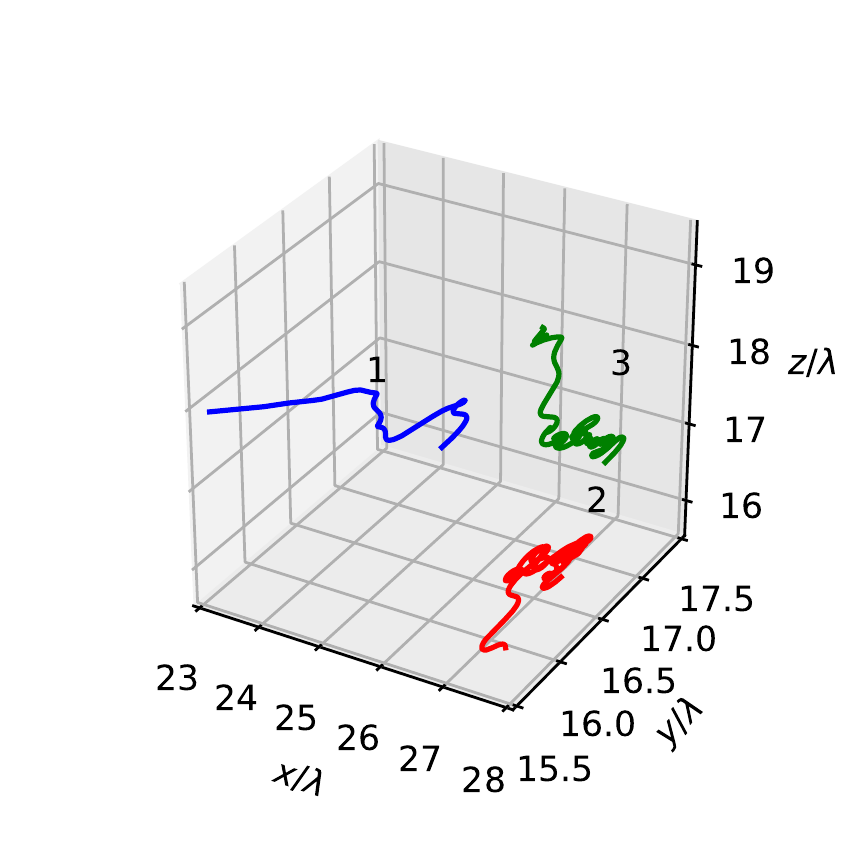} 
\caption{The trajectories of the escaping  inner-shell ($2s^{1}$)
electrons of Xe (line 1) and the inner-shell electrons of Xe (lines 2 and 3) staying for
a long time in the region, where the laser field peaks.}
\label{Xe-track} 
\end{figure}

In contrast to He the electrons are still produced by
field ionization of Xe even after laser pulse crossing ($t\geq18\lambda/c$)
because the ionization potential of the inner-shell electrons of Xe
atom is in about $3$ order of magnitude higher than that of He. Typical
trajectories of the $2s^{1}$ electrons of Xe are shown in Fig.~\ref{Xe-track}.
It is seen that some inner-shell electrons escape from the high-intensity
region while the other inner-shell electrons undergo oscillations
in the strong laser field for a long time and emit high-energy photons.
It follows from Fig.~\ref{Xe-distr} that like for He the
density of the electrons, positrons an ions also peak in the nodes
of the standing wave forming near volume center.

\begin{figure}[h]
 \includegraphics[width=8cm]{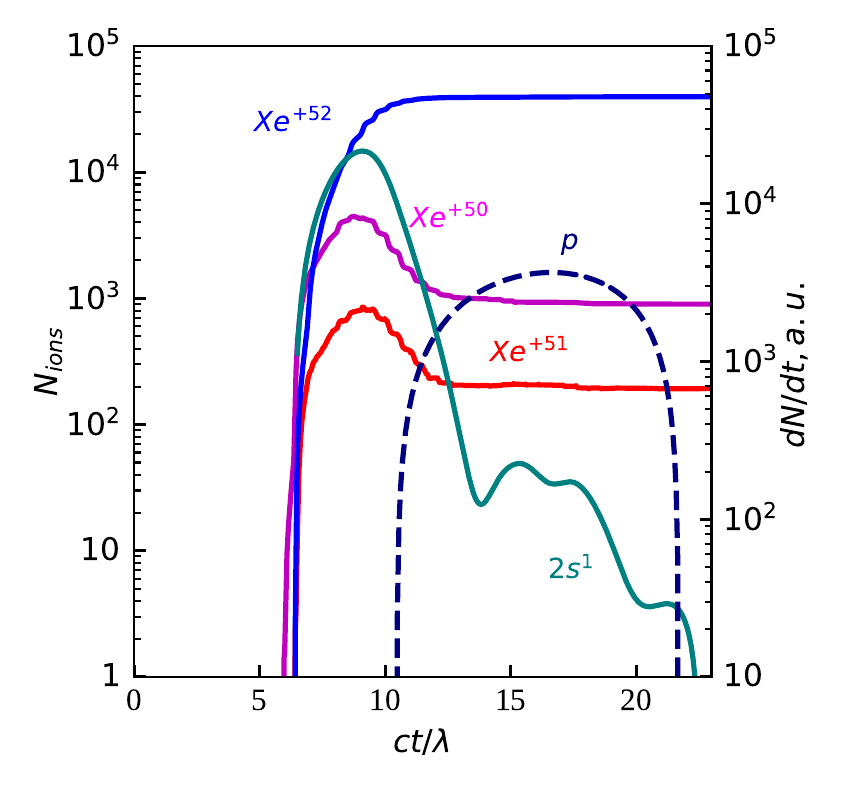} 
\caption{The number of the high-charge ions, the ionization production
rate for the inner-shell electrons and the positron production rate
as a function of time  in Xe. }
\label{ions} 
\end{figure}

The dynamics of the inner-shell electron population and of the high-charge
ion population is shown in Fig.~\ref{ions}. It is seen from Fig.~\ref{ions}
that the ions with the highest charges Xe $^{+52} $ are produced when
the laser pulses crosses and the laser field strength peaks. The ionization
production rate for the inner-shell electrons and the positron production
rate are significant at $ 7\lambda/c < t<22\lambda/c$ when the the laser field is strong. 
Therefore, the cascade in Xe can be initiated not only by
the outer-shell electrons trapped in the standing wave but also by
the inner-shell electrons produced when the laser field becomes to
be strong enough for cascade development. 

\begin{figure}[h]
\centering \includegraphics[width=8cm]{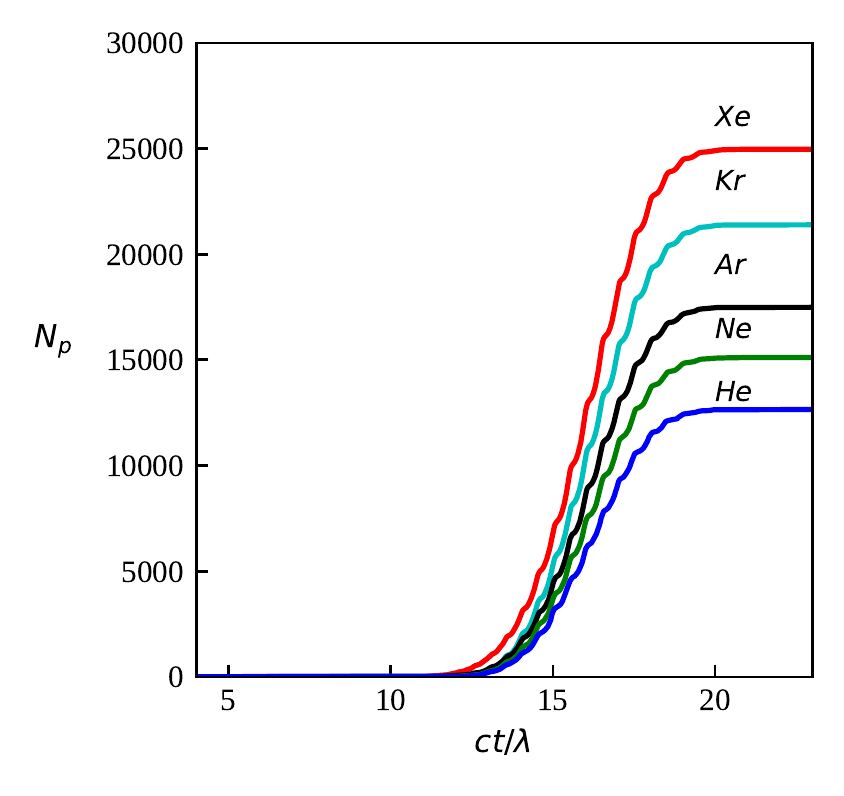} 
\caption{The pair number as a function of time for  He, Ne, Ar, Kr and Xe. }
\label{pair} 
\end{figure}

\begin{figure}[h]
\centering \includegraphics[width=8cm]{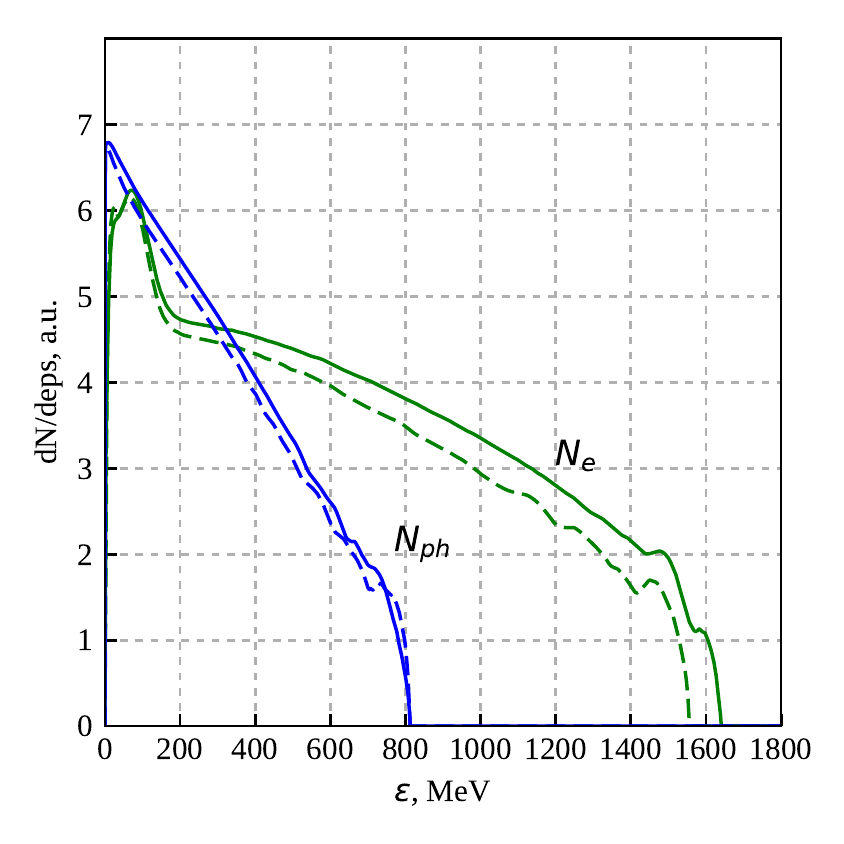} 
\caption{The spectrum of the photons in He (dashed blue line) and in Xe (solid
blue line) and the spectrum of the electrons in He (dashed green line)
and in Xe (solid green line) at $t=28\lambda/c$. }
\label{spectra} 
\end{figure}

The pair numbers as function of time in all noble gases are shown in Fig.~\ref{pair}.
The gas densities are normalized to the atomic numbers so that the electron densities in the case of  full atom ionization are the same for all gases. The gas densities of He, Ne, Ar, Kr and Xe are $9.03\times10^{15}\mbox{c}\mbox{m}^{-3}$, $1.81\times10^{15}\mbox{c}\mbox{m}^{-3}$,
$10^{15}\mbox{c}\mbox{m}^{-3}$, $5.02\times10^{14}\mbox{c}\mbox{m}^{-3}$
and $3.34\times10^{14}\mbox{c}\mbox{m}^{-3}$, respectively. The gas volume has a length of $5 \lambda/c $ and a width of $5 \lambda /c $. It is seen from Fig.~\ref{len} that ratio $N_p(Xe)/ N_p(He) $ keeps nearly unchanged with increasing of volume size. Thus, we can expect that the ratio in the positron numbers presented in Fig.~\ref{pair} for the noble gases will be similar for a macroscopic gas target ($L \gg 40 \lambda /c$). Despite
the fact that Xe density is in $27$ times less than He
density the number of pairs produced in Xe is in about $2$ times
larger than that in He. Therefore the inner-shell electrons play
an important role in QED cascade triggering. The spectra of the
electrons and photons produced in the cascade  in He and Xe
are shown in Fig.~\ref{spectra}. The energy of the particles is
slightly higher in Xe than in He.

\section{Discussions and conclusions}

The simple formula for the field ionization rate covering all range of laser
intensity is proposed. The formula based on combination of known expression for tunnel ionization in the low intensity limit  \cite{Perelomov1966-1,Popov2004,Karnakov2015,Ammosov1986} and the ionization rate formula in the extremely intense limit where
the rate is proportional to the strength of the electric field. The linear dependence on the field strength is in a qualitative
agreement with numerical TDSE calculations for hydrogen \cite{Bauer1999} if $E \gg E_{cr}$.
However more detailed validation of the proposed formula is needed. 

QED cascades in noble gases are studied. It is shown that there are
two main mechanisms of seed electron production and cascade initiation in high-Z gases like
Ar, Kr and Xe: (i) the ionization of the outer-shell electrons
moving along with the pulses to the cascade region and (ii)  the ionization
of the inner-shell electrons created at the instance when the pulses crosses
and the total laser field peaks. The ionization potential of $2s^{1}\text{and }2s^{2}$
electrons of Xe are about $10$ keV. Those electrons can escape from
the ion only at very high field strength. In low-Z gases like
He and Ne only the first mechanism is possible. These gases
are fully ionized in the laser pulse front. Most of the electrons
are pushed out by the ponderomotive force of the laser pulse from the high
intensity region and cannot initiate cascade. The small part of the
produced electrons may move along with the laser pulses up to the
time instance when the pulses crosses and the laser field becomes
strong enough for cascading. 

The first mechanism is discussed in Ref.~\cite{Jirka2016} where electron trapping in the cascade region
is observed in numerical simulations. It is demonstrated that part of the seed electrons survives up to pulse crossing and is trapped near the electric field nodes of the standing wave formed by the counter-propagating pulses. The second mechanism are discussed in Ref.~\cite{Tamburini2016} where ionization is included in cascade simulations. However the ionization model was too simple and does not take into account probabilistic nature of ionization and sequential multiple ionization of electrons from different shells of high-Z atoms. Moreover all atom electrons leave the atom simultaneously according to this model. We use in our simulations more realistic model providing probabilistic description of field ionization. This model allows us analyse the role  of both mechanisms. It follows from our simulations that in high-Z gases like Ar, Kr and Xe both mechanisms are important for cascade initialization: the outer-shell electrons are involved in the first mechanism while the inner-shell electrons are involved in the second one. Comparing pair production in He and Xe (see Figs.~\ref{len} and \ref{pair}) we can conclude that the inner-shell electrons of Xe increase pair production in about $2$ times for the parameters of interest despite the fact that Xe density is in $27$ times less than He
density. The result is obtained when the peak laser field strength $ 2 a =1000 $ is close the cascade threshold value. With increasing of the laser intensity the role of the mechanisms can be changed.

QED cascade develops as a result of chain reactions when photon emission
caused by electron scattering in the laser field (Compton scattering)
alternates with pair photoproduction due to photon scattering in the
laser field (Breit-Wheeler process). However high-energy photons and
pairs can be also produced by collisional processes neglected in our
numerical simulations. Photons can be emitted at electron scattering
by ionic or atomic nuclei (bremsstrahlung) while pairs can be created
by a high-energy photon interacting with ionic or atomic nuclei. First
we estimate the bremsstrahlung contribution to the cascade. The
total bremsstrahlung cross-section in the limit $\varepsilon_{e}\gg\varepsilon_{e}^{\prime}\gg m_e c_{e}^{2}$
is $\sigma_{br}=\left(2^{5/2}/3\right)Z^{2}\alpha\cdot r_{e}^{2}\gamma_{e}^{1/2}$,
where $\varepsilon_{e}$ and $\varepsilon_{e}^{\prime}$ are the electron
energies before and after scattering by an ion, respectively, $\gamma_{e}=\varepsilon_{e}/\left(m_{e}c^{2}\right)$
is the relativistic gamma-factor of the electron,  $\alpha=e^{2}/\left(c\hbar\right)\approx1/137$
is the fine structure constant, $r_{e}=e^{2}/\left(m_{e}c^{2}\right)\simeq2.82\times10^{-13}$~cm
is the classic radius of the electron \cite{Landau4}. The number
of the bremsstrahlung photons can be estimated as follows $N_{bph}\simeq n_{e}N_{i}c\sigma_{br}\tau_{c}$,
where $n_{e}$ is the density of the relativistic electrons, $N_{i}$
is the number of the the ions in the cascade volume and $\tau_{c}$
is the cascade duration. According to simulation the cascade volume
is $V_{c}\sim5\lambda\times5\lambda\times5\lambda$ and the cascade
duration is $\tau_{c}\simeq7\lambda/c$ (see Fig.~\ref{ions}). For
full ionization of xenon gas with $Z=54$ and $n_{g}=3.34\times10^{14}$~cm$^{-3}$
we obtain $n_{e}=Zn_{g}\simeq1.8\times10^{16}$~cm $^{-3}$, $N_{i}=n_{g}V_{c}\simeq4.18\times10^{4}$
and $N_{bph}\simeq9.51\times10^{-5}$ where the mean gamma-factor
of the electron $\gamma_{e}\simeq10^{3}$ is used. It follows from
estimation that the number of the bremsstrahlung photons is negligible
with number of the cascade photons ($N_{ph}>10^{5}$). 

The number
of the electron-positron pairs produced by the high-energy photons
near nuclei can be estimated as follows $N_{npair}\simeq n_{ph}N_{i}c\sigma_{npair}\tau_{c}$,
where $\sigma_{br}\simeq\left(28/9\right)Z^{2}\alpha\cdot r_{e}^{2}\left[\ln\left(2\varepsilon_{ph}/m_e c^{2}\right)-\left(109/42\right)-1.2\left(\alpha Z\right)^{2}\right]$
is the cross-section of pair production near nuclei in the relativistic
limit ($\varepsilon_{e,ph}\gg m_e c^{2}$) and $n_{ph}=N_{ph}/V_{c}$
is the photon density in the cascade volume \cite{Landau4}. It follows
from the simulations that the number of the high-energy photons ($\varepsilon_{ph}>1$~MeV)
is less than $N_{ph}<10^{6}$ therefore the number of pairs produced
by photons near nuclei is $ N_{npair}<10^{-5}$ that is much less than
the pair number produced in the cascade $N_{pair}>10^{4}$.

The contribution of the collisional ionization can be also estimated
by similar way. The cross-section of the collisional ionization for
the relativistic electrons is $\sigma_{ci}\simeq\left(2^{7/2}\pi^{1/2}/3\right)Z^{2}\alpha\cdot r_{e}^{2}L$,
where $L$ is the Coulomb logarithm \cite{Wu2017}. Even for very
large value of $L=20$ the number of electrons produced via collisional
ionization $N_{e,ci}=n_{e}N_{i}c\sigma_{ci}\tau_{c}\simeq0.016$ is
much less than the number of the electrons created via field ionization
$N_{e}\gg N_{i}>10^{4}$ in the cascade volume. Thus collision ionization
can be neglected in the cascade modeling. 

The electrons can be also
produced by the collision of high-energy photons with atoms or partially
ionized ions (photoelectric effect). This effect of one-photon ionization
is not included in our numerical scheme for field ionization. The
photoelectric cross-section peaks for photons with energy $\varepsilon_{ph}<m_e c^{2}$
and does not exceed $\sigma_{phe}<10^{-19}$cm $^{-2}$ \cite{Storm1970}.
It follows from Fig.~\ref{spectra} that the mean energy of the cascade
photons is more than $100$~MeV and the photons with energy $\varepsilon_{ph}<m_e c^{2}$
belong the low-energy part of the photon spectra. The number of the
photons radiated by the electron moving in the laser field per unit
time can be estimated as follows $dN_{ph}/d\varepsilon_{ph}=\varepsilon_{ph}^{-1}dI/d\varepsilon_{ph}$
and $dN_{ph}/d\varepsilon_{ph}\simeq\omega_{L}^{-1}0.021\alpha a_{S}\chi_{e}^{2/3}\gamma_{e}^{-4/3}\left(\varepsilon_{ph}/m_e c^{2}\right)^{-2/3}$,
where the approximation for the low-energy part of the synchrotron
radiation spectrum is used, $dI/d\varepsilon_{ph}$ is the synchrotron radiation spectrum \cite{Baier1998} and $a_{S}=m_e c^{2}/\hbar\omega_{L}$.
The number of the low-energy photons emitted during cascade development
is $N_{ph}\left(\varepsilon_{ph}<m_e c^{2}\right)\simeq0.063\alpha a_{S}\chi_{e}^{2/3}\gamma_{e}^{-4/3}\left(\omega_{L}\tau_{c}\right)\simeq4$,
where $ \chi_{e}\simeq10$ is taken from the simulations. The number
of the electron produced via photoelectric effect in the cascade region
during cascade development is $N_{e,phe}=\left(N_{ph}/V_{c}\right)N_{i}c\sigma_{phe}\tau_{c}\simeq10^{-7}$
that is much less than the number of the electrons created via field
ionization $N_{e}\gg N_{i}>10^{4}$. Therefore photoelectric effect
can be also neglected in the cascade modeling. The losses associated
with the ionization are also neglected in the simulations because
the ionization energy ($ <0.1m_e c^{2}$) is in several order of magnitude
less than the mean electron energy in the laser field ($\sim10^{3}m_e c^{2}$).

Finally it is demonstrated the Xe among noble gases is more appropriate to facilitate QED cascading. However the additional effects like the laser pulse propagation from the focusing parabolas to the cascade volume, the accurate description of the gas target and ionization dynamics should be taken into account for realistic simulations of possible laboratory experiments.

\begin{acknowledgments} 

This work was supported by part by the "Basis" Foundation Grant No. 17-11-101-1. The numerical simulations of QED cascades was supported by the Russian Science Foundation Grant No. 16-12-10383. 

\end{acknowledgments}

\end{document}